\documentclass[aip,pof,reprint,onecolumn]{revtex4-1}
\usepackage{graphicx}
\usepackage{subfigure}

\usepackage{latexsym,amsfonts,amssymb}

\begin{document}

\title{Poiseuille flow past a nanoscale cylinder in a slit channel: Lubrication theory versus molecular dynamics analysis}

\author{Amir M. Rahmani}

\author{Yang Shao}

\author{Mehlam Jupiterwala}

\author{Carlos E. Colosqui}
\email[]{carlos.colosqui@stonybrook.edu}
\affiliation{Department of Mechanical Engineering, Stony Brook University, Stony Brook, NY 11794, USA.}


\begin{abstract}
Plane Poiseuille flow past a nanoscale cylinder that is arbitrarily confined (i.e., symmetrically or asymmetrically confined) in a slit channel is studied via hydrodynamic lubrication theory and molecular dynamics simulations, considering cases where the cylinder remains static or undergoes thermal motion.
Lubrication theory predictions for the drag force and volumetric flow rate are in close agreement with molecular dynamics simulations of flows having molecularly thin lubrication gaps, despite the presence of significant structural forces induced by the crystalline structure of the modeled solid.  
While the maximum drag force is observed in symmetric confinement, i.e., when the cylinder is equidistant from both channel walls, the drag decays significantly as the cylinder moves away from the channel centerline and approaches a wall. 
Hence, significant reductions in the mean drag force on the cylinder and hydraulic resistance of the channel can be observed when thermal motion induces random off-center displacements.
Analytical expressions and numerical results in this work provide useful insights into the hydrodynamics of colloidal solids and macromolecules in confinement. 
\end{abstract}


\maketitle 

%
%
\section{Introduction}
\label{sec:introduction}
%
Understanding hydrodynamic phenomena at the nanoscale has become increasingly important with the advent of numerous new technologies enabled by nanofabrication methods \cite{gates2005,bocquet2010}.
Predicting hydrodynamic forces and volumetric rates in nanoscale flows is particularly relevant to the design of nano-electromechanical systems (NEMS), nanowire nanosensors, and nanofluidic devices for applications that range from bioengineering to materials science and renewable energy \cite{hong2003,mijatovic2005,patolsky2005,song2008,ekinci2010,JFM2010}.
At length scales in the order of one nanometer, which corresponds roughly to the size of three water molecules, fundamental assumptions upon which classical hydrodynamic equations are derived need to be thoroughly evaluated.
For example, experimental studies indicate that solid-like transitions and viscoelastic behavior can arise for simple liquids confined in nanoscale lubrication gaps \cite{israelachvili1989,gee1990,klein1995}.
Transitions to viscoelastic behavior of simple fluids are found to occur when hydrodynamic time scales are comparable to the relaxation time of the fluid, which in micro/\-nanoscale confinement can exceed by several orders of magnitude its bulk value \cite{hu1991,POF2009}. 
Another fundamental issue for continuum-based descriptions of micro/\-nanoscale flows is the difficulty to determine proper boundary conditions (e.g., no slip, Navier/\-Maxwell slip) that are determined by physicochemical properties of the fluid and the nanoscale structure of the confining solid surfaces \cite{neto2005,lauga2007,colosqui2013slip}.
In this context, molecular dynamics (MD) simulations have become a valuable tool to study hydrodynamic phenomena in nanoscale confinement and to help in developing and validating continuum-based descriptions for nanoscale flows.
Previous works have demonstrated the use of different MD techniques (e.g., equilibrium/\-non-equilibrium) to determine hydrodynamic forces, boundary conditions, and transport coefficients resulting from complex interfacial phenomena in nanoscale confinement \cite{koplik1995,thompson1992,koplik1988,travis2000,cieplak2001,leng2005,leng2006}.
Along similar lines, the present work resorts to fully atomistic non-equilibrium MD simulations in order to develop (continuum-based) hydrodynamic descriptions that yield quantitative predictions for nanoscale lubrication flows past micro/\-nanoscale bodies that are perfectly static or subject to thermal motion.

Hydrodynamic lubrication theory is suitable to study flows within lubrication films having nonuniform thickness and arbitrary shape.    
Hydrodynamic models for nanoscale films, however, can encounter significant limitations due to complex interfacial phenomena that arise near fluid-solid interfaces.
For example, nanoscale roughness of the confining solid surfaces can produce lubrication forces resulting from both liquid-solid and solid-solid friction. 
Moreover, layering of liquid atoms at the surface of a crystalline solid and the dynamic rearrangement of the induced solid-like structures can lead to strong structural forces and so-called ``stick-slip'' motion\cite{thompson1990,reiter1994,bhushan1995,urbakh2004,lei2011}.
Numerous experimental studies using a surface force apparatus or atomic force microscope have probed and characterized lubrication forces in molecularly-thin films at different shear rates \cite{luengo1996,carpick1997,kavehpour2004,ruths2008,bonaccurso2008,israelachvili2010}.
Notably, in the case of liquids with simple molecular structure (e.g., molecular chains such as n-alkanes) confined by molecularly smooth surfaces, tribological studies report that the shear viscosity within the lubrication film does not vary significantly with respect to the liquid bulk value for shear rates as high as 10$^5$ 1/s and lubrication gaps as small as ten molecular 
diameters \cite{israelachvili1989,ruths2011,al2013effects}.  
Furthermore, these studies indicate that the slip plane lies within one molecular diameter from the solid surface independently of the presence of electrostatic double-layers or structural forces.
For the particular case of water confined by smooth silica surfaces the same conclusions hold for lubrication gaps as small as 2 nm (i.e., about the size of six water molecules) \cite{israelachvili1989,horn1989,kuhl1998,raviv2004}. 
Hence, experimental evidence indicates that hydrodynamic lubrication theory is able to yield reliable predictions for atomically smooth surfaces and molecularly-thin lubrication gaps (i.e., as thin as five to ten molecular layers) under a wide range of flow conditions (e.g., Couette-type flows with moderate-to-high shear rates).  
In this work we study creeping flow of a simple molecular liquid past a colloidal solid cylinder confined in a slit channel and lying at arbitrary distances from the channel centerline. 
In Sec.~II we first we study the case where the cylinder is perfectly static, employing hydrodynamic lubrication theory we obtain analytical expressions for the drag forces and flow rates in finite channels.  
The studied problem involves two lubrication gaps of variable height that can become molecularly thin as the cylinder approaches contact with a channel wall.
In Sec.~III we describe the MD technique employed and simulations performed to provide a microscopic description of nanoscale flows, without relying on conventional continuum assumptions.
In Sec.~IV we assess the validity of the hydrodynamic lubrication approach for the case of static cylinders of micro/nanoscale dimensions. 
Drag forces and volumetric flow rates predicted for symmetric and asymmetric confinement, in channels with different lengths, are compared against numerical solution of the Navier-Stokes (N-S) equations and fully atomistic MD simulations.
In Sec.~V, employing predictions obtained for static conditions we study the case of a colloidal cylinder that performs random displacements induced by thermal motion.
The approach in this section indicates ways in which hydrodynamic lubrication models can be applied to predict mean drag forces and flow rates for confined nanoscale bodies (e.g., nanoparticles, macromoleucles, nanowires, nanobeams) that are subject to thermal motion.
%
%
%
%
\section{Poiseuille flow past a static cylinder arbitrarily confined}
\label{sec:introduction}
\begin{figure}[h]
\vspace{-0pt}
\begin{center}
\includegraphics[width=0.65\textwidth]{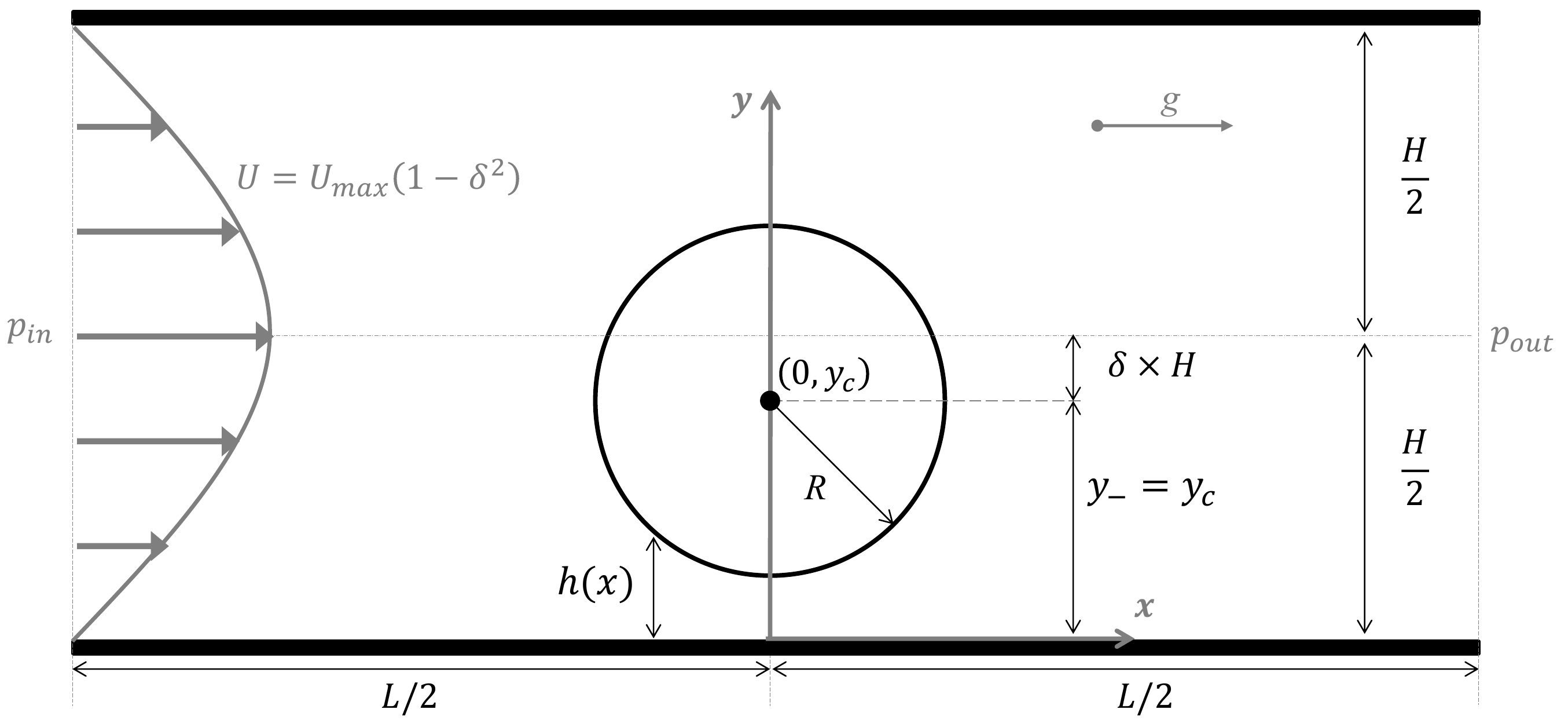}
\end{center}
\vspace{-0pt}
\caption{Plane Poiseuille flow past a confined cylinder with arbitrary off-center displacement ($|\delta|\leq(1-k)/2$) and high confinement ratio ($k=2R/H>0.5$).
Flow at a constant volumetric rate is driven by a pressure differential ($\Delta p=p_{in}-p_{out}$) and/or a constant body force ($\rho g$) in the $x$-direction.
The channel is considered to be sufficiently long to develop a parabolic velocity profile $U(y)=U_{max} (1-\delta^2)$ at the inlet and outlet ($x=\pm L/2$).
\label{fig:1}}
\vspace{-0pt}
\end{figure}

The geometry of the studied flow problem is illustrated in Fig.~\ref{fig:1}, a circular cylinder of radius $R$ is fully confined within a slit channel of height $H$, width $W \gg H$, and length $L\gg H$.
Under studied conditions the flow is assumed to be steady, two-dimensional, incompressible, isothermal, and Newtonian; the fluid density $\rho$ and shear viscosity $\mu$ are thus assumed constant.
The cylinder center is located at ($x=0,y=y_c$) and thus lies at a vertical distance $\delta \times H=H/2-y_c$ from the channel centerline (cf. Fig~\ref{fig:1}).
To characterize the studied flow we will employ the confinement ratio
\begin{equation}
\label{eq:confinement_ratio}
k=\frac{2R}{H}
\end{equation}
and the dimensionless off-center displacement, or asymmetry parameter,  
\begin{equation}
\label{eq:asymmetry_parameter}
\delta=\frac{1}{2}-\frac{y_c}{H}.
\end{equation}
In addition, the dimensionless channel length $l=L/H$ will be employed to characterize finite length effects on the volumetric flow rate and drag force for long but finite channels ($l\gg1$).
%
%

%
%
%
%
%
%
Flow in the $x$-direction at constant volumetric flow rate (per unit width) $Q$ [m$^2$/s] is sustained by a driving force $G/W=(p_{in}-p_{out}) H + \rho g (H L -\pi R^2)$; here, $p_{in}$ and $p_{out}$ are the static pressures at the channel inlet and outlet, respectively, and $\rho g$ is a constant body force active on the fluid phase.
The channel is assumed to be sufficiently long so that a parabolic velocity profile $u(\pm L/2,y)=U_{max} (1-\delta^2)$ with $U_{max}=3Q/2H$ is established at the channel inlet and outlet.
%
The studied conditions correspond to creeping flows with very small Reynolds numbers 
$Re=\rho U_{max} H/\mu \ll 1$.
The drag coefficient is thus defined as
\begin{equation}
\label{eq:drag_coefficient}
\lambda(k,\delta)=\frac{D}{\mu U}, 
\end{equation}
where $D$ is the drag force per unit width and $U=u(-L/2,y_c)$ is the velocity of the ``unperturbed'' flow velocity upstream of the cylinder center. 

For the case of symmetric confinement ($\delta=0$) and moderate confinement ratio $0.2\lesssim k\lesssim 0.5$ there are available expressions for the drag coefficient obtained from approximate solution of the Stokes equations with no-slip boundary conditions:
\begin{equation}
\lambda(k,0)=\frac{4\pi}{A_0-(1+0.5k^2+A_4k^4+A_6 k^6+A_8k^8)\ln\,k+B_2k^2+B_4k^4+B_6k^6+B_8k^8}
\label{eq:faxen}
\end{equation} 
with $A_0=0.9156892732$, $A_4= 0.05464866$, $A_6=−0.26462967$, $A_8= 0.792986$, $B_2= 1.26653975$, $B_4=−0.9180433$, $B_6= 1.877101$, and $B_8=−4.66549$ as derived by Fax\'{e}n \cite{faxen1946}.
A similar expression for symmetric confinement and $k\lesssim0.5$ has been derived via analytical solution of the Oseen equations by Takaisi \cite{takaisi1956}. 
For the case of asymmetric confinement ($\delta\ne0$), which has received considerably less attention, a perturbative solution of the biharmonic equation performed by Harper\cite{harper1967} has provided an approximate analytical expression for $\lambda(k,\delta)$ that is only valid for $k\ll1$ (i.e., for $R\ll H$).
This work is primarily concerned with flow configurations with arbitrary off-center displacements of the cylinder, $0\leq|\delta|\leq(1-k)/2$, and high confinement ratio, $k \to 1$, where a lubrication flow approximation is valid.  
%
%
%
%
%
%
\subsection{Hydrodynamic Lubrication Theory \label{sec:lubrication}}
For the prediction of drag forces and flow rates via hydrodynamic lubrication theory we will assume Newtonian flow regimes and no slip boundary conditions. 
Results from numerical solution of the full Navier-Stokes (NS) equations and MD simulations for different configurations ($k\ge0.5$, $\delta \ge 0$, $l>5$) will be compared against the derived analytical expressions.
Results from fully atomistic MD simulations will assess the validity of the adopted assumptions in the case of nanoscale flows of simple molecular liquids confined by wettable surfaces that are atomically smooth.
The local height along the channel is $h_{\pm}(x)=H$ for $|x|\ge R$ and 
\begin{equation}
h_{\pm}(x)=\left(\frac{1}{2}\pm\delta\right)H-\sqrt{R^2-x^2}~~\mathrm{for}~~|x|\le R, 
\label{eq:h}
\end{equation}
where the $(+)$ and $(-)$ signs correspond to the lubrication gaps above and below the cylinder, respectively.
While for $\delta=0$ both lubrication gaps are equal, either gap fully closes for $|\delta| = (1-k)/2$. 
In clearing the cylinder the flow rate $Q$ splits into $Q_-=\alpha(k,\delta) Q$, flowing below the cylinder, and $Q_+=(1-\alpha)Q$ flowing above the cylinder.
The split factor $\alpha$ can be determined by equating the pressure drops $\Delta p_-=\Delta p_+=p(R)-p(-R)$ across the bottom and top lubrication gaps, which yields the following equation:
\begin{equation}
\alpha(k,\delta)\int_{-R}^R \frac{dx}{h^{3}_{-}}=[1- \alpha(k,\delta)] \int_{-R}^R \frac{dx}{h^{3}_{+}}.
\label{eq:alfa-dp}
\end{equation}
While the split factor in Eq.~\ref{eq:alfa-dp} takes the expected value $\alpha(k,0)=1/2$ in symmetric confinement, $\alpha(k,\delta)\to 1$ for $\delta \to (k-1)/2$ (i.e., when closing the top gap) and $\alpha(k,\delta)\to 0$ for $\delta \to (1-k)/2$ (i.e., when closing the bottom gap).

After establishing the flow rates $Q_\pm$ via Eq.~\ref{eq:alfa-dp}, it is straightforward to determine the drag coefficient $\lambda(k,\delta)$ using conventional lubrication analysis.
The full derivation of the drag coefficient is presented in the Appendix, while the main analytical results are summarized in this section.
The drag coefficient predicted via lubrication theory is
\begin{equation}
\lambda(k,\delta)=\frac{8}{(1-\delta^2)}\{\alpha [f_p(k,\delta)-f_s(k,\delta)]-(1-\alpha)f_s(k,-\delta)\},
\label{eq:lambda}
\end{equation}
with the flow split factor given by the explicit expression 
\begin{equation}
\alpha(k,\delta)=\frac{1}{1+f_p(k,\delta)/f_p(k,-\delta)}.
\label{eq:alpha}
\end{equation}
The shape functions in Eqs.~\ref{eq:lambda}--\ref{eq:alpha} are
\begin{equation}
\label{eq:fp}
f_p(k,\delta)=\frac{3}{4}\frac{k^2(1/2-\delta)}{b^{5/2}}\left[\frac{\pi}{2}+\mathrm{atan}\left(\frac{k}{2b^{1/2}}\right)\right]+\frac{3k^3}{8b^2}\frac{1}{(1/2-\delta)}+\frac{1}{(1/2-\delta)}\frac{k}{b},
\end{equation}
and  
\begin{equation}
\label{eq:fs}
f_s(k,\delta)=\frac{k^2}{4b^{3/2}}\left[\frac{\pi}{2}+\mathrm{atan}\left(\frac{k}{2b^{1/2}}\right)\right]+\frac{k}{2b}.
\end{equation}
Here, the confinement parameter $b=(y_c^2-R^2)/H^2\equiv (1/2-\delta)^2-k^2/4$ is introduced for a more compact definition of the shape functions $f_p$ and $f_s$ accounting for pressure and shear drag contributions, respectively.

In the limit $k\to 1$ the dominant contribution in Eq.~\ref{eq:lambda} is due to pressure forces that are proportional to the lubrication parameter $\epsilon^{-\frac{5}{2}}$, here $\epsilon=(H-2R)/2R=(1-k)/k$ is the nondimensional effective gap height.
Hence, Eq.~\ref{eq:lambda} predicts two limit cases:
\begin{equation}
\lambda(k\to 1,0)=\frac{12\pi}{\sqrt{2}}\epsilon^{-\frac{5}{2}}
\label{eq:highk0}
\end{equation}
for cases of symmetric confinement $\delta=0$; and  
\begin{equation}
\lambda(k\to 1,\delta_{max})
=3\pi\epsilon^{-\frac{5}{2}}
\label{eq:highkd}
\end{equation}
for the maximum cylinder displacement $|\delta_{max}|=(k-1)/2$ where one of the lubrication gap is fully closed.
Eq.~\ref{eq:lambda} recovers the asymptotic behavior $\lambda\propto\epsilon^{-\frac{5}{2}}$ in lubrication flows as $\epsilon\to 0$ \cite{jeffrey1981,stone2005,ben2004}.
Notably, $\lambda(k\to 1,\delta_{max})=\lambda(k\to1,0)/(2\sqrt{2})$ and for high confinement ratios there is significant reduction in the drag coefficient as the cylinder approaches contact with the top or bottom channel wall. 

Depending on the particular application, either the flow rate $Q$ or induced the driving force $G$ (i.e., pressure differential and body forces) is prescribed.
When the flow rate is prescribed knowing the drag coefficient suffices to predict the drag force $D=\mu U \lambda$ where $U=(3Q/2H)(1-\delta^2)$.
When the driving force is prescribed, however, it is necessary to predict the volumetric flow rate (per unit width) $Q(k,\delta,l)$ in order to predict the drag force and the hydraulic resistance for different confinement configurations and channel aspect ratios.
The lubrication flow approximation yields 
\begin{equation}
Q(k,\delta,l)=\frac{(p_{in}-p_{out})/L + \rho g}{12\mu/H^3} \phi(k,\delta,l)
\label{eq:Q}
\end{equation}
determined by the flow correction factor
\begin{equation}
\phi(k,\delta,l)=\frac{l}{l +\alpha(k,\delta) f_p(k,\delta) -k},
\label{eq:phi}
\end{equation}
where $l=L/H$ is the dimensionless channel length, and $\alpha$ and $f_p$ are given by Eqs.~\ref{eq:alpha}--\ref{eq:fp}.
%
According to Eqs.~\ref{eq:Q}--\ref{eq:phi}, the volumetric rate $Q_\infty$ for Poiseuille flow is recovered for $l \to \infty$ and the flow rate vanishes $Q\to0$ for $k\to1$.  
It is worth noticing that for long but finite channel lengths ($l\gg 1$) the flow rate increases as the cylinder is displaced from the channel centerline ($\delta >0$) according to the ratio
\begin{equation}
\frac{Q(k,\delta,l)}{Q(k,0,l)}=  \frac{l +f_p(k,0)/2-k}{l +\alpha(k,\delta) f_p(k,\delta) -k}.
\label{eq:Qratio}
\end{equation}

A few comments are in order about the expressions derived for the drag coefficient and drag force.
For the particular case of symmetrically confined cylinders, predictions from Eqs.~\ref{eq:lambda}--\ref{eq:fs} for the drag coefficient $\lambda(k,0)$ are in close quantitative agreement with asymptotic formulas for $k\to 1$ proposed in previous work \cite{ben2004}. 
For asymmetrically confined cylinders, Eq.~\ref{eq:lambda} predicts a significant decrease in the drag coefficient $\lambda(k,\delta)$ as $|\delta| \to (1-k)/2$. 
Moreover, the derived formulas predict a maximum drag force $D$ for symmetric confinement and significant drag reduction in asymmetric confinement with reduction ratios that depend on the dimensionless channel length $l=L/H$. 
To the best of our knowledge an analytical expressions analogous to Eq.~\ref{eq:lambda} and Eq.~\ref{eq:Q}, valid for cylinders in Poiseuille-type flows for high confinement ratio ($k\gtrsim 0.5$) and arbitrary off-center displacement ($0\le|\delta|\le (k-1)/2$), are not available in the previous literature.
%

%
%
%
\section{Molecular Dynamics Simulation \label{sec:md}}
Following standard techniques for non-equilibrium MD simulations \cite{allen1990,frenkel2002}, the interaction between any two atoms of species $s$ and $s'$ is governed by a generalized Lennard-Jones (LJ) potential 
\begin{equation}
\label{eq:LJ}
U_{LJ}^{s,s'}(r_{ij})=4\epsilon\left[\left(\frac{\sigma}{r_{ij}}\right)^{12}-A_{ss'}\left(\frac{\sigma}{r_{ij}}\right)^6\right]. 
\end{equation}
Here, $r_{ij}=|{\bf r_i}-{\bf r_j}|$ is the separation between any two atoms ($i,j=1,N$), $\sigma$ is the repulsive core diameter, and $\epsilon$ is the depth of energy potential minimum, which lies at $r_{ij}=(2/A_{ss'})^{1/6}\sigma$. 
%
%
%
%
%
\begin{figure}[h!]
\vspace{-0pt}
\begin{center}
\includegraphics[width=0.6\textwidth]{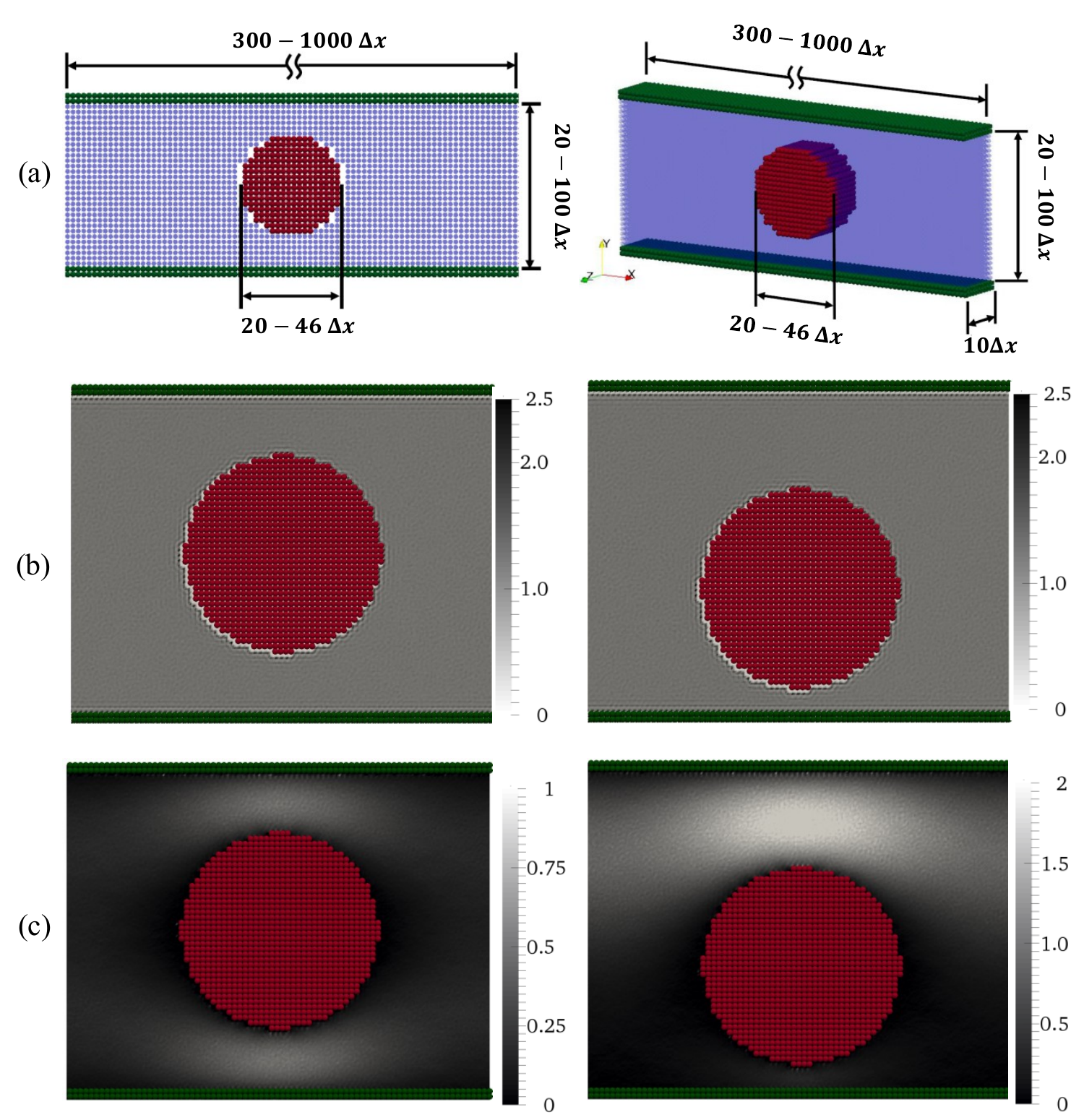}
\end{center}
\vspace{-0pt}
\caption{Geometric setup and flow features in MD simulations. 
(a) Side and perspective views of the simulation domain and range of dimensions employed; the side view shows the initial fcc atomic lattice with constant spacing $\Delta x=0.8^{-1/3}\sigma$.
(b) Dimensionless mass density $\rho/\overline{\rho}$ where $\overline{\rho}=0.8 m/\sigma^3$ is the mean bulk density; solid-like structure and layering of fluid atoms is observed near the solid surfaces. 
(c) Dimensionless momentum density magnitude $\rho|\bf{u}|/\overline{\rho}|\bf{u}_{0}|$ where $\bf{u}_{0}$ is the maximum velocity in the symmetric confinement case ($\delta=0$). As the bottom gap closes $\delta \to (1-k)/2=0.15$ the flow through the upper gap becomes twice the value observed in symmetric confinement.
Reported quantities in (b--c) are obtained via time average and spatial average in the $z$-direction, for $k=0.7$ and $l=5$ in cases of symmetric confinement $\delta=0$ (left panels) and asymmetric confinement at $\delta=0.11$ (right panels).
\label{fig:MDsetup}}
\end{figure}
The simulated system is composed of three atomic species that correspond to the fluid $(s=1)$, cylindrical particle $(s=2)$, and channel walls $(s=3)$ (cf. Fig.~\ref{fig:MDsetup}).
The symmetric attraction coefficients $A_{ss'}=A_{s's}$ control the degree of wettability of the modeled solid surfaces and the shear-dependent hydrodynamic slip length; the parametrization employed ($A_{12}=A_{13}=0.8$) produces highly wettable solids exhibiting very small hydrodynamic slip on flat or curved surfaces over a wide range of flow conditions \cite{drazer2002,drazer2005,colosqui2013,razavi2014}.
For the simulations in this work, fluid atoms conform dimer molecules bound by Finitely Extensible Nonlinear Elastic (FENE) potentials
\begin{equation}
\label{eq:FENE}
U_{FENE}(r_{ij})=-\frac{1}{2} k_F r_{max}^2 \log\left[ 1- \left(\frac{r_{ij}}{r_{max}}\right)^2 \right], 
\end{equation}
where $k_F$ is the stiffness of the modeled molecular bond and $r_{max}$ adjust its maximum extension.
The use of FENE potentials in addition to LJ interactions allows further control of rheological properties and the volatility of the modeled fluid.
A Nose-Hoover thermostat maintains constant fluid and solid atoms at temperature $T={\epsilon}/{k_B}$ ($k_B$ is the Boltzmann constant).
At initialization the atoms of all species are arranged in face-centered cubic (fcc) lattice with constant spacing $\Delta x=n^{-1/3}$ (see Fig.~\ref{fig:MDsetup}a), where $n=0.8/\sigma^3$ was the number density employed in all MD simulations in this work.
The mean mass density of the fluid $\overline{\rho}=0.8 m/\sigma^3$ is constant (here $m$ is the atomic mass), for the modeled conditions the shear viscosity is $\mu=2.9\sqrt{m \epsilon}/\sigma^2$.

Non-equilibrium MD simulations are performed to study drag forces and volumetric rates for Poiseuille-type flow past a nanoscale cylindrical particle confined at arbitrary off-center displacements $\delta$. 
The net force on the cylindrical particle $(s=2)$ is ${\bf F}=- \sum \partial U^{1,2}_{LJ}/\partial {\bf x}$ obtained as the sum of all atomic interactions with the fluid $(s=1)$ only; i.e., direct atomic interactions between the cylindrical particle and the channel walls are neglected in our MD simulations.
Different nanochannels with heights $H=26$--$90\Delta x$ and lengths $L=300$--$1000\Delta x$ are employed in MD simulations in order to characterize the drag and flow rates in a range of confinement ratios ($k=$ 0.6--0.84) and channel aspect ratios ($l \simeq$5--38); a constant width $H=10\Delta x$ is employed in all cases.
We consider the solid-liquid interface to be located at the zero isopotential contour $U_{LJ}=0$ for the solid species (i.e., particle and walls); the confinement ratio $k$ and dimensionless channel length $l$ in MD simulations were calculated following this criterion.
A constant body force $\textbf{f}= m g \textbf{i}$ is applied to each fluid atom in order to drive the flow in the $x$-direction and periodic boundary conditions are applied in the $x$ and $z$ directions (cf. Fig.~\ref{fig:MDsetup}). 
In all cases the applied driving force produces flows with low Reynolds numbers $Re=\rho U_{max} H/\mu \le 0.3$, where $U_{max}=\rho g H^2/8 \mu$. 
Distinctive features of the mean mass and momentum density fields simulated via MD are reported in Figs.~\ref{fig:MDsetup}(b--c). 
Near the liquid-solid interface we observe layering of fluid atoms near the solid surfaces and a small but finite amount of hydrodynamic slip that varies locally.
As the cylinder approaches contact with a channel wall (cf. Fig.~\ref{fig:MDsetup}c) the flow through the narrower lubrication gap rapidly decreases as quantitatively predicted by Eq.~\ref{eq:alpha}.
%
%
%
%
%
\section{Static Cylinders}
Predictions from lubrication theory are compared against numerical simulations via finite element solution of the steady-state N-S equations \footnote{The commercial package COMSOL was employed for numerical solution of the Navier-Stokes equations in this work.} and MD techniques described in Sec.~\ref{sec:md}.
In our numerical simulations periodic boundary conditions are applied at the channel inlet and outlet, a constant body force in the $x$-direction results in a total force magnitude $G=\rho g (HL-\pi R^2) W$ driving the flow.  
%
%
%
%
%
\begin{figure}[h]
\vspace{-0pt}
\begin{center}
\begin{center}
\includegraphics[width=0.8\textwidth]{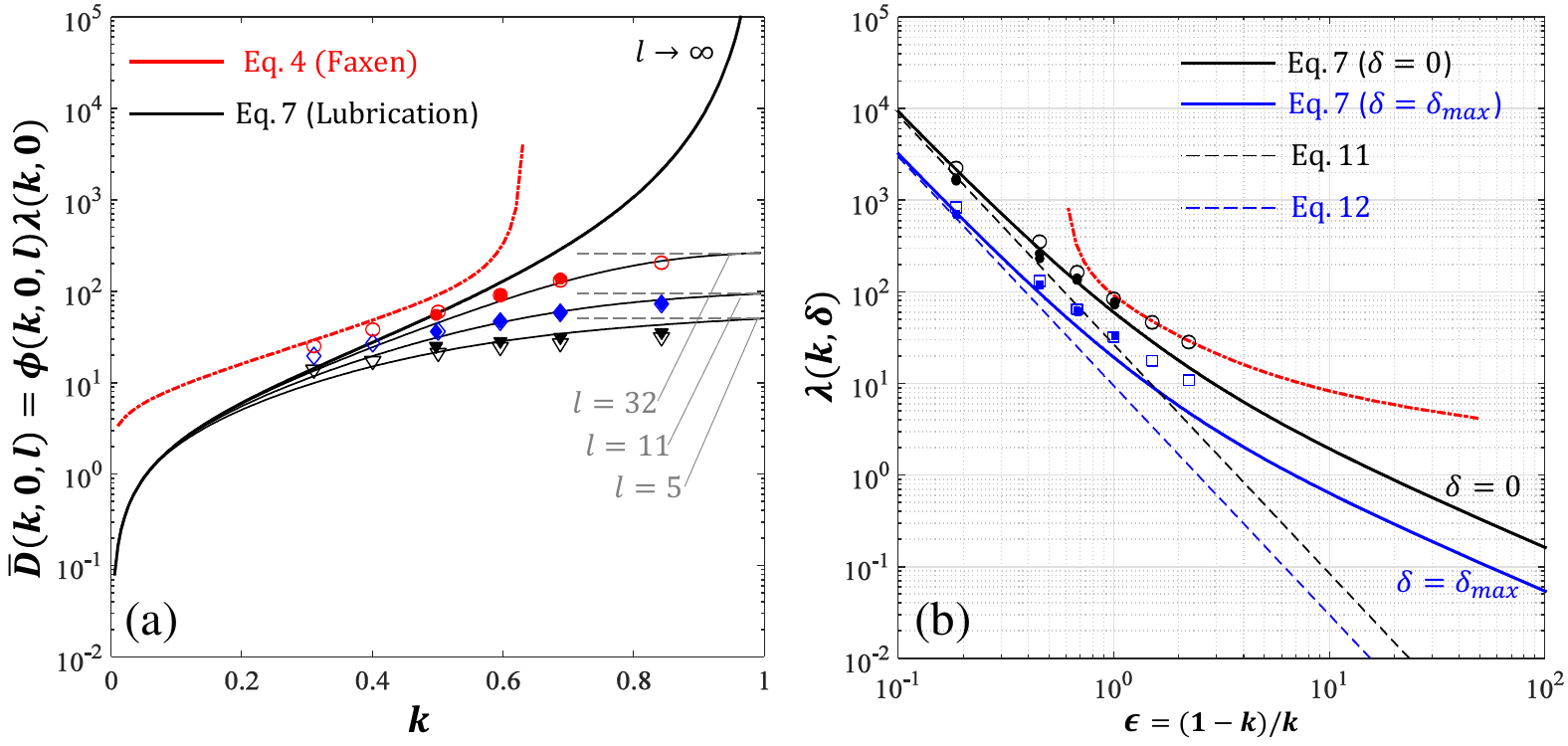}
\end{center}
\end{center}
\vspace{-0pt}
\caption{Drag forces and drag coefficients; theoretical predictions (solid/dashed lines), N-S simulation (open markers), and MD simulations (filled markers).
(a) Normalized drag force $\bar{D}(k,\delta,l)=D/\mu (3Q_\infty/2H)=\phi \lambda$ for symmetrically confined cylinders ($\delta=0$) versus confinement ratio $k=2R/H$ for $l=L/H=$~5--32.
The flow correction factor $\phi(k,\delta,l)$ is given by Eq.~\ref{eq:phi}.
Plotted for comparison (in both panels) are predictions for infinitely long channels ($l\to\infty$ and $\phi=1$) based on Faxen's formula (Eq.~\ref{eq:faxen}) and lubrication theory (Eq.~\ref{eq:lambda}) [see legend].
For a finite driving force $G$ in the limit $k\to1$ where $Q\to0$, the drag force becomes equal to the driving force $D=G$ and thus $\bar{D}\to 8 l$ (dashed lines).
(b) Drag coefficient $\lambda(k,\delta)=D/\mu U$ where $U=(3Q/2H)(1-\delta^2)$ as a function of $\epsilon=(H-2R)/2R=(1-k)/k$, for $\delta=0$ (i.e., symmetric confinement) and $|\delta|=(1-k)/2$ (i.e., limit case of asymmetric confinement where the cylinder contacts either channel wall).
For $\epsilon\to 0$ the drag coefficient exhibits the asymptotic behavior predicted in Eqs.~\ref{eq:highk0}--\ref{eq:highkd} with significant drag reduction in asymmetric confinement.
\label{fig:dragforce}}
\vspace{-0pt}
\end{figure}
%
%
%
%
\begin{figure}[h!]
\begin{center}
\includegraphics[width=0.8\textwidth]{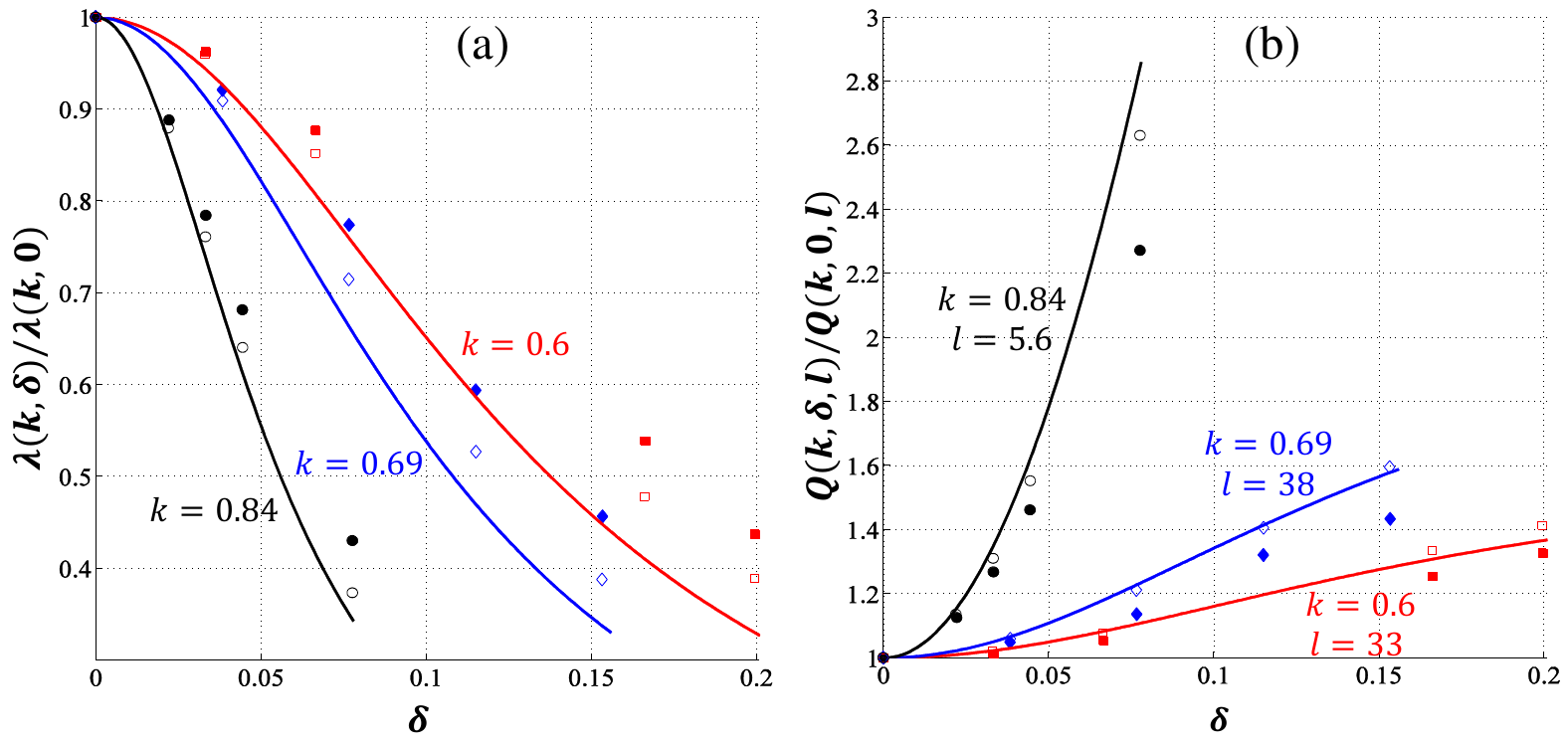}
\end{center}
%
\caption{Drag reduction and flow enhancement (i.e., hydraulic resistance reduction) in asymmetric confinement; theoretical predictions (solid lines), N-S simulation (open markers), and MD simulations (filled markers).
(a) Drag coefficient ratio $\lambda(k,\delta)/\lambda(k,0)$ as a function of the dimensionless off-center displacement $\delta$ for three different confinement ratios $k$.
Significant drag reduction is observed in asymmetric confinement $|\delta|>0$ as the confinement ratio increases ($k\to1$); as expected, agreement between simulations and lubrication theory predictions from Eq.~\ref{eq:lambda} increases for $k\to 1$. 
(b) Flow rate enhancement $Q(k,\delta,l)/Q(k,0,l)$ as a function of the dimensionless off-center displacement $\delta$ for for three different finite channels [see legend]. 
The hydraulic resistance of the channels increases when the confined cylinder moves away from the center.
\label{fig:lambda}}
\end{figure}
While physical conditions modeled in (continuum-based) N-S simulations correspond to macroscopic channels where $H/\sigma \gg 1$, the conditions modeled in MD simulations correspond to channels with nanoscale dimensions $H/\sigma=$~20--90 (i.e., $H \simeq$~5--30~nm).   
The reported drag forces $D$ are obtained by subtracting the cylinder weight and buoyancy force from the total force computed in numerical simulations and the unperturbed velocity $U=(3Q/2H)(1-\delta^2)$ is determined from the numerically computed flow rate (per unit width) $Q$.
In the case of MD simulations, reported quantities correspond to averages over sufficiently long times $T_a>0.2 H L/Q$ for which convergence of the reported mean values (within a 10\% deviation) is observed after reaching a steady flow rate.

We first analyze the results for the hydrodynamic drag force $D$ on a cylinder in symmetric confinement conditions ($\delta=0$).
The predicted drag force is $D(k,0,l)= \mu (3Q_\infty/2H) \phi(k,0,l) \lambda(k,0)$ where $\lambda$ is given by Eq.~\ref{eq:lambda} and $\phi$ is given by Eq.~\ref{eq:phi}; 
here, $Q_\infty=G H^2/12 \mu L W$ is the flow rate expected for an infinitely long channel ($l \to \infty$) for the finite driving force $G$ applied in numerical simulations. 
As showed in Fig.~\ref{fig:dragforce}a, theoretical predictions for the drag force as a function of the confinement ratio $k$ in symmetric confinement conditions are in close agreement with both numerical solution of the N-S equations and MD simulations for long channels with dimensionless length $l=L/H=$~5--32.   
For $k\to 1$, as both (bottom/top) lubrication gaps close and the flow vanishes ($Q\to 0$) the force on the cylinder balances the applied driving force $D(k\to 1, \delta, l)=G$ (cf. \ref{fig:dragforce}).
It is worth noticing (cf., Fig.~\ref{fig:dragforce}a) that for moderate confinement ratios ($k\simeq$~0.5--0.7) as the dimensionless channel length increases ($l>10$) the hydrodynamic drag force on the cylinder becomes less than half the force applied to drive the flow ($D/G<0.5$).
%

The drag coefficient $\lambda(k,\delta)=D/\mu U$ predicted by lubrication theory (Eq.~\ref{eq:lambda}) is compared against numerical simulations in Fig.~\ref{fig:dragforce}b for the case of symmetric confinement, where $\delta=0$, and the limit case when the cylinder contacts either one of the channel walls, where $|\delta|=(1-k)/2$.
Results from N-S and MD simulation confirm the expected asymptotic behavior $\lambda\propto\epsilon^{-5/2}$, where $\epsilon= (H-2R)/2R$, for $k\to1$ that is predicted by Eqs.~\ref{eq:highk0}--\ref{eq:highkd}.
In the limit case $|\delta|=(1-k)/2$ where the cylinder contacts one of the channel walls there is a reduction of about 65\% with respect to the drag coefficient in symmetric confinement (cf. Fig.~\ref{fig:lambda}b).
Numerical solution of the N-S equations and MD simulations confirm a gradual reduction in the drag coefficient as the dimensionless off-center displacement $\delta$ increases (cf. Fig.~\ref{fig:lambda}a,).
The drag coefficient ratio $\lambda(k,\delta)/\lambda(k,0)$ quantifying the reduction of drag in asymmetric confinement as a function of $\delta$ is reported in  Fig.~\ref{fig:lambda}b for three different confinement ratios $k=0.6, 0.69, 0.84$.
As expected the agreement between lubrication theory and numerical simulations improves for large confinement ratios ($k\gtrsim0.7$).
Notably, drag coefficients computed from MD simulations of flows having molecularly thin lubrication gaps (i.e., three to ten atomic layers thin) are in good agreement with numerical solutions of the full Navier-Stokes equations and analytical predictions from lubrication theory adopting no-slip boundary conditions (cf. Fig.~\ref{fig:lambda}a).  
The lubrication analysis in Sec.~\ref{sec:lubrication} also predicts a gradual reduction in the hydraulic resistance of the channel as the off-center displacement of the confined cylinder increases.
This effect is observed in simulations as an enhancement in the flow rate $Q$ for a prescribed driving force $G$ as the dimensionless off-center displacement $\delta$ increases.
As showed in Fig~\ref{fig:lambda}b, theoretical predictions from Eq.~\ref{eq:Qratio} for the flow enhancement ratio $Q(k,\delta,l)/Q(k,0,l)$ are in close agreement with numerical simulations for different confinement ratios ($k=$~0.6, 0.69, 0.84) and finite channels with different lengths ($l=$~5.6, 33, 38). 
Simulations confirm predictions of Eq.~\ref{eq:Qratio}, the flow enhancement for asymmetrically confined cylinders increases for large confinement ratios ($k\to1$) and decreases for long channels ($l\to\infty$). 

\section{Colloidal cylinders}
The lubrication analysis in Sec.~\ref{sec:lubrication} produced analytical predictions for the position-dependent drag force assuming flow past a perfectly static cylinder. 
This section discusses the application of hydrodynamic lubrication to confined colloidal cylinders that undergo Brownian motion and can be 
strongly influenced by colloidal interactions (e.g., van der Waals attraction, steric repulsion, oscillatory structural forces). 
The employed approach is generally applicable to colloidal particles, whether these are freely convected or bound to an equilibrium position by diverse restoring forces. 
The formulas presented in this section, via invoking analytical predictions from Sec.~\ref{sec:lubrication}, aim to predict the mean (noise-averaged) drag force experienced by nanobeams, nanowires, or colloidal probes that constitute a key component of NEMS and nanowire-based sensors and actuators.

For overdamped Brownian motion (i.e., neglecting inertial and memory effects), the mean drag force $\langle D \rangle$ on a confined colloidal particle can be estimated by ensemble averaging the (position-dependent) drag for static conditions over the sequence of random displacements induced by thermal motion.
Since the drag predicted under static conditions varies only in the vertical direction ($y$-direction), uncorrelated thermal motion in the flow direction ($x$-direction) is not expected to affect the mean drag force.
Vertical random displacements can be statistically described by a probability density $\varrho_o(\delta,t)\equiv\varrho(\delta,t| \delta_0,t_0)$; here $\delta_0$ is the initial displacement at time $t_0$ where $\varrho_o(\delta,t_0)=\delta(\delta-\delta_0)$.
Hence, the mean drag force expected for overdamped Brownian motion is 
\begin{equation}
\langle D(k,l,t) \rangle= \frac{3\mu}{2H} \int_{-\delta_{max}}^{+\delta_{max}} Q(k,\delta,l) \lambda(k,\delta) \varrho_o(\delta,t) d\delta,  
\label{eq:Dmean}
\end{equation} 
where $\lambda(k,\delta)$ is the drag coefficient (Eq.~\ref{eq:lambda}) derived for static conditions and $\delta_{max}=(1-k)/2$ as before.
Similarly, it is useful to define a noise-averaged drag coefficient
\begin{equation}
\langle \lambda(k,t) \rangle
= \int_{-\delta_{max}}^{+\delta_{max}} \lambda(k,\delta) \varrho_o(\delta,t) d\delta,
\label{eq:Lmean}
\end{equation} 
in order to characterize the mean drag force $\langle D(k,l,t) \rangle$ for the case where the flow rate $Q$ is prescribed; this case also corresponds to prescribing the driving force in sufficiently long channels ($l\to\infty$) for which the flow correction factor (Eq.~\ref{eq:phi}) becomes unity $\phi(k,\delta,l\to\infty) = 1$.

Via MD simulations we analyze the case of a confined colloidal cylinder immersed in a fluid with constant thermal energy $k_B T$. 
A linear restorative force $F_s=K_s \sqrt{(x-L/2)^2+(y-H/2)^2}$ is applied to bring the colloidal cylinder to equilibrium at the center of the channel where $\delta=0$;
the ``spring'' constant is varied in the range $K_s$~=~0--1~$k_B T/\sigma^2$ in order to modulate the root-mean-square (rms) amplitude of the dimensionless off-center displacement 
$\delta_{rms}(t)=\sqrt{\langle(\delta(t)-\delta_0)^2\rangle}$.    
The mean drag and drag coefficient defined in Eqs.~\ref{eq:Dmean}--\ref{eq:Lmean} are expected to depend on the dimensionless rms displacement $\delta_{rms}=\delta_{rms}$ observed in MD simulations.
In order to produce analytical predictions we will assume that the colloidal cylinder follows an Ornstein-Uhlenbeck process and thus 
\begin{equation}
\varrho_o(\delta,t)=\frac{1}{\sqrt{2\pi}\delta_{rms}(t)} \exp\left\{-\left[\frac{\delta-\delta_0 \exp(t/\tau)}{\sqrt{2}\delta_{rms}(t)}\right]^2\right\},
\label{eq:gaussian}
\end{equation}
where $t_0=0$ and $\tau=\tilde{D}/K_s$ is an unknown effective diffusion coefficient.
Since boundary effects are neglected, the probability in Eq.~\ref{eq:gaussian} is only valid for small rms displacements $\delta_{rms}(t)\ll\delta_{max}=(1-k)/2$.
For finite values $K_s$ there is long-time limit $t>>\tau$ where initial condition is forgotten and the probability in Eq.~\ref{eq:gaussian} becomes $\varrho(\delta,t)=\exp[-(\delta/\sqrt{2}\delta_{rms})^2]/\sqrt{2\pi}\delta_{rms}$.

%
%
\begin{figure}[h]
\begin{center}
\includegraphics[width=0.85\textwidth]{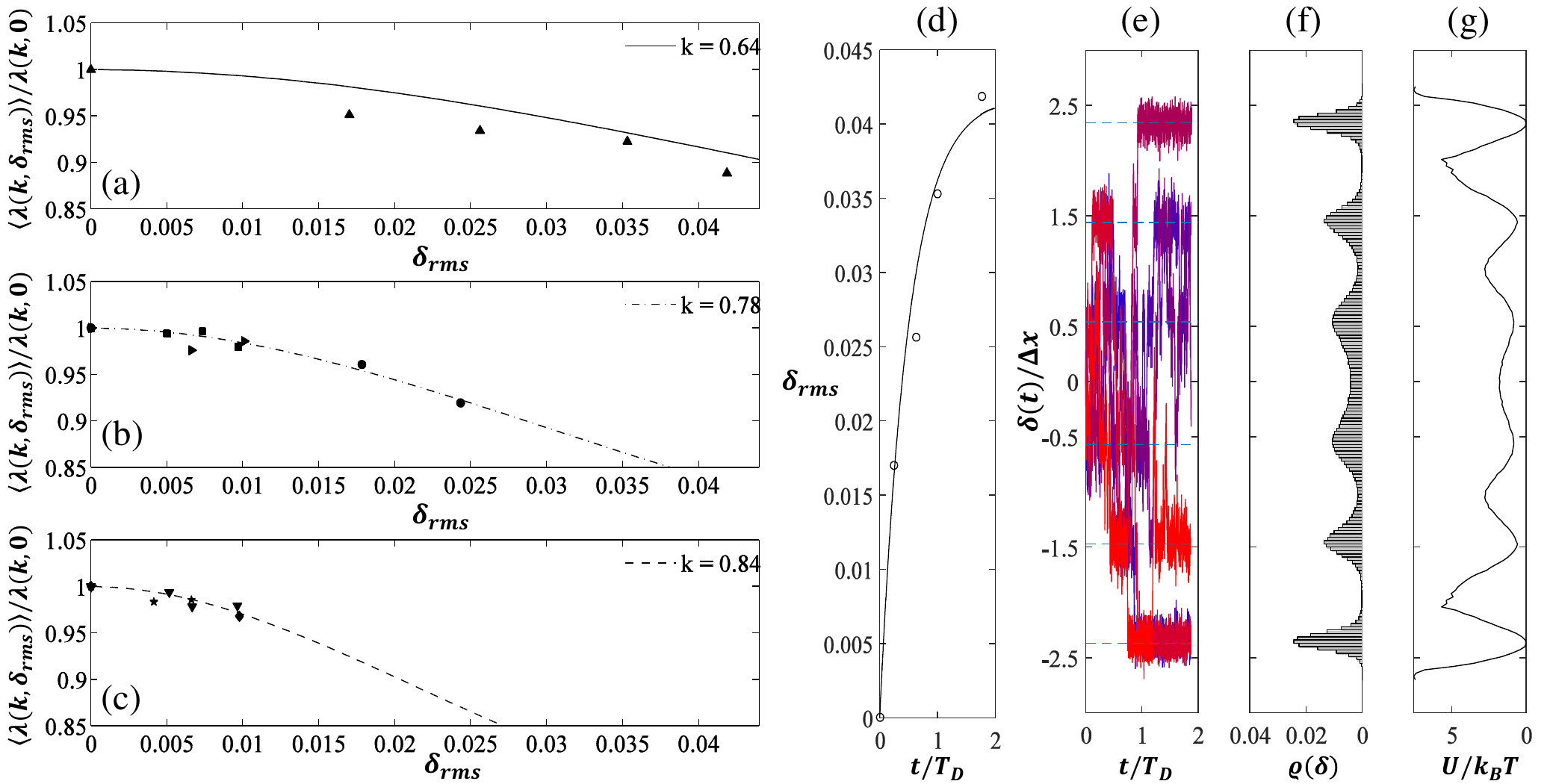}
\end{center}
%
\caption{Drag reduction induced by thermal motion and structural forces for a colloidal cylinder symmetrically confined in a slit channel. 
(a--c) Mean drag coefficient reduction $\langle\lambda(k,\delta_{rms})\rangle/\lambda(k,\delta_{rms})$ versus rms off-center displacement.
Plotted lines indicate predictions from Eq.~\ref{eq:Lmean} for $k=0.64,0.78, 0.84$. 
Filled markers correspond to results from MD simulations: 
(a) ($\blacktriangle$) $k=0.64$, $l=3.56$ ($x_{rms}$=0); 
(b) ($\bullet$) $k=0.74$, $l=3.5$, ($\blacktriangleright$) $k=0.74$, $l=3.5$ ($x_{rms}$=0), ($\blacksquare$) $k=0.78$, $l=5$; 
(c) ($\blacklozenge$) $k=0.82$, $l=5$, ($\blacktriangledown$) $k=0.84,$ $l=5$ ($x_{rms}$=0), ($\bigstar$) $k=0.86$, $l=3$.  
(d) Time evolution of rms displacement versus dimensionless time $t/T_D$ ($T_D=\mu k H^3/2k_B T$) for MD simulations in panel (a); solid line is an exponential fit, open markers correspond to values computed from MD simulation.   
(e) Vertical off-center displacement in lattice units $\Delta x$ versus dimensionless time $t/T_D$ for MD simulations in panel (a); six different realizations showing metastable states.
(f) Stationary probability distribution $\varrho(\delta)$ computed from absolute value of displacement-time trace in panel (e) via ensemble average over six realizations. 
(g) Free energy $U(\delta)$ computed from probability distribution in panel (f).
\label{fig:thermal}
}
\end{figure}

Predictions for the mean drag coefficient (Eq.~\ref{eq:Lmean}) via adopting the probability density in Eq.~\ref{eq:gaussian} for $t>>\tau$ and $\delta_{rms}\ll\delta_{max}$ are compared in Fig.~\ref{fig:thermal} against results from MD simulations for colloidal cylinders symmetrically confined in nanoscale channels of various heights $H=$~30--90$\sigma$. 
For the MD simulations reported in Fig.~\ref{fig:thermal}a the cylinder is allowed to ``freely'' drift in the vertical direction while the motion is prescribed in the $x$-direction; the expected rms vertical displacement is $\delta_{rms}=2\tilde{D}t$ (for $t>\tau$) and the (top/bottom) lubrication gaps in these MD simulations become as small as two atomic diameters. 
For the MD simulations reported in Figs.~\ref{fig:thermal}b--c a restorative force with different strengths ($K_s$~=~0.1--1~$k_B T/\sigma^2$) is applied; in this case the maximum rms displacement is bounded, $\delta_{rms}=\sqrt{k_B T/K_s}$, and lubrication gaps in this case are always larger than five atomic diameters.
In all cases, MD simulations (cf. Fig.~\ref{fig:thermal}) report a decay in the mean drag force as the rms displacement $\delta_{rms}$ increases, in close agreement with predictions from Eq.~\ref{eq:Lmean}.
However, the rms displacements reported in Figs.~\ref{fig:thermal}a--c as computed from MD simulations can become significantly smaller than the expected values for diffusion in homogeneous fluid media when the cylinder reaches within three atomic layers from the channel walls.
Moreover, for the case where the cylinder is free to drift vertically we observe a slow exponential relaxation determined by a diffusive time $\tau\sim T_D=\mu k H^3/2k_B T$ (see Fig.~\ref{fig:thermal}d). 
In fact, when a lubrication gap becomes thinner than five atomic layers the displacement-time trace of the cylinder center-of-mass exhibits long-lived metastable states (cf. Fig.~\ref{fig:thermal}d) that indicate the local energy minima that lie at average separation $\Delta x=(\rho/m)^{-1/3}$ (cf. Fig.~\ref{fig:thermal}e).

Assuming Boltzmann statistics, the stationary probability distribution ($t\to\infty$) is $\varrho(\delta)=Z^{-1} \exp[-U(\delta)/k_B T]$, where $U(\delta)$ is the (space-dependent) free energy and $Z$ is the proper normalization constant.  
A strongly non-Gaussian probability distribution computed from the displacement-time trace reveals an oscillatory free energy $U(\delta)=-k_B T\log(\varrho) + const.$ that decays away from the walls.  
This observation explains the poorer agreement observed in Fig.~\ref{fig:thermal}a between MD simulations and predictions adopting a Gaussian probability in Eq.~\ref{eq:gaussian} valid for free Brownian motion in the long-time limit $t\gg T_D$.  
Given that our MD simulations do not include atomic interactions between solid atoms, the oscillatory free energy variations are attributed to the structural rearrangement of fluid layers caused by the cylinder motion.
Hence, the modeled steric and van der Waals interactions between solid and fluid atoms induced significant energy barriers ($\Delta U=5 k_B T$) and long-lived metastable states when the cylinder is close to the wall.
%

%
%
\section{Conclusions}
A hydrodynamic lubrication approach was presented to predict drag forces and volumetric rates for plane Poiseuille flow past a confined static cylinder as a function of the confinement ratio $k$, the dimensionless off-center displacement of the cylinder $\delta$, and the dimensionless channel length $l$.
Analytical expressions for the drag coefficient introduced in this work are valid for moderate to large confinement ratios ($k\gtrsim 0.5$) and arbitrary off-center displacements
 ($0\le|\delta|\le(1-k)/2$). 
In the high confinement limit $k\to 1$ the derived expressions recover the asymptotic behavior reported in previous works\cite{ben2004,semin2009}.
The set of derived formulas applies to cases when either the volumetric flow rate or the driving force is prescribed.
In addition, the derived expressions valid for finite channels are suitable for predicting drag forces and volumetric rates for flow past periodic arrays of cylinders.

As the cylinder moves away from the channel centerline and one of the lubrication gap closes, either above or below the cylinder, the flow through the closing gap vanishes and so does its contribution to the drag force.
The derived expressions quantitatively predict that (i) drag forces and drag coefficients have their maximum value in symmetric confinement ($\delta=0$), and (ii) there are significant reductions in both the drag force and drag coefficient in asymmetric confinement, as the cylinder approaches either one of the channel walls ($|\delta| \to 1/2-k/2$).
Conversely, the hydraulic resistance for a given confinement ratio $k$ is minimized when the cylinder contacts a wall and $|\delta|=1/2-k/2$.

In the case of static cylinders, analytical predictions for the drag force and flow rates are in good agreement with numerical solutions of the Navier-Stokes equations and fully-atomistic MD simulations of nanoscale channels.  
Notably, conventional hydrodynamic descriptions adopting no-slip boundary conditions produced reliable predictions despite the presence of significant steric effects and structural forces observed in fully atomistic simulations with wettable solids.
The results in this work indicate that hydrodynamic lubrication theory can produce reasonable predictions for molecularly thin lubrication gaps (i.e. down to three atomic layers) in the case of Poiseuille-type flows on plane channels with surfaces that are molecularly smooth and highly wettable by simple molecular liquids.
In fact, MD simulations and continuum models (i.e., lubrication theory and numerical solution of the N-S equations) reported comparable values of the drag force when one of the lubrication gaps became vanishingly small.
The observed agreement, however, can be attributed to the fact that the flow through the narrowest lubrication gap decreases, as quantitatively predicted by equating the pressure drop through each gap, and the dominant contribution to the drag force comes from the widest lubrication gap, which in all studied cases remained thicker than two atomic layers.  
The MD simulations reported a small hydrodynamic slip that depended on the local surface curvature and shear rate magnitude but this effect did not affect significantly the agreement with analytical predictions adopting no slip boundary conditions.
The presented lubrication analysis can be readily extended to systems with partially wettable solids where significant hydrodynamic slip is present, provided that the slip length is a known parameter. 

The studied case of nanoscale cylinders undergoing thermal motion revealed a few important effects. 
For symmetrically confined colloidal cylinders the mean (noise-average) drag force, determined via ensemble or time average, can be significantly lower than the drag force predicted for a static cylinder. 
The mechanism for the predicted drag reduction is not attributed to hydrodynamic slip but rather to the colloidal cylinder randomly moving to off-center positions where the drag predicted in static conditions is significantly lower.
Similar thermally-induced effects produce a noticeable reduction in the mean hydraulic resistance as the rms displacement increases.
For creeping flows and after assuming a Gaussian probability density for the thermally-induced displacements of the colloidal cylinder, the reduction in the mean drag and hydraulic resistance can be quantitatively predicted by averaging the position-dependent drag and flow rate derived in static conditions.
The observed effects can be enhanced by increasing the rms amplitude of the cylinder displacement via different mechanisms, which can include mechanical or acoustic actuation and/or increasing the fluid temperature.
Under studied conditions, the presence of significant structural forces was found to be the major obstacle to safely extending hydrodynamic lubrication theory to nanoscale flows in plane channels. 
Analytical or numerical solution of a Fokker-Planck equation can predict the probability density of random thermal displacements but this will require a priori knowledge of local free energy variations for a confined colloidal cylinder.
Although structural forces did not play a significant role when the cylinder position was prescribed, oscillatory structural forces induced strongly non-Gaussian probability densities and long-lived metastable positions of the cylinder at integer number of atomic layers from the channel wall.
The analysis and results presented in this work are relevant to the design of NEMS and nanowire-based sensors and actuators, nanofluidic devices for transport and separation of nanoparticles or macromolecules, and can potentially guide experimental studies of the nanorheology of confined fluids using colloidal probes.

\begin{acknowledgments}
The authors would like to thank Antonio Checco, Joel Koplik, and Yongsheng Leng for useful discussions.
This work was supported by the SEED Grant Program by The Office of Brookhaven National Laboratory (BNL) Affairs at Stony Brook University. 
Part of the MD simulations in this work employed computational resources from the Center for Functional Nanomaterials at BNL, which is supported by the U.S. Department of Energy, Office of Basic Energy Sciences, under Contract No. DE-SC0012704.
\end{acknowledgments}

\appendix
\section{Drag force and flow rate derivation via lubrication theory}
The steady-state, incompressible, and isothermal Navier-Stokes equations in the lubrication limit are reduced to:
\begin{equation}
\mu\frac{\partial^2 u}{\partial y^2}-\frac{\partial p}{\partial x}+\rho g=0; ~~~\frac{\partial p}{\partial y}=0.
\label{eq:NSL}  
\end{equation}
Here, $\mu$ is the shear viscosity, $\rho$ is the fluid velocity, and $g$ is acceleration due to a constant body force.
Solution of Eq.~\ref{eq:NSL} gives the velocity profile
\begin{equation}
u(x,y)=\frac{1}{2\mu}\left(\frac{\partial p}{\partial x}-\rho g\right)\left(y^2-h(x)y\right)
\end{equation}
and the volumetric flow rate (per unit width)
\begin{equation}
Q=\frac{h^3(x)}{12\mu}\left(\frac{\partial p}{\partial x}-\rho g\right).
\end{equation}
As showed in Fig.~\ref{fig:1}, the local height is $h(x)=H$ for $|x|>R$ and $h(x)=h_{\pm}(x)$ for $|x|\leq R$; here $h_{\pm}(x)=\left(\frac{1}{2}\pm\delta\right)H-\sqrt{R^2-x^2}$ where the $(+)$ and $(-)$ signs correspond to the top and bottom lubrication gaps, respectively.
In clearing the cylinder the flow must split into $Q_-=\alpha(k,\delta)Q$ below the cylinder and $Q_+=\alpha(k,-\delta)Q$ above the cylinder; the split factor $\alpha$ must satisfy mass conservation and thus $\alpha(k,-\delta)=1-\alpha(k,\delta)$. 
The pressure gradient inside the bottom lubrication gap is given by
\begin{equation}
\label{pressure_diff}
\frac{\partial p_-}{\partial x}=\rho g - 12\mu \alpha(k,\delta) Q\frac{1}{h_{-}^3}
\end{equation}
Exact analytical integration of Eq.~\ref{pressure_diff} gives the pressure drop across the bottom gap
\begin{equation}
\label{pressure_drop}
\Delta p_-= p(R)-p(-R)=\rho g 2R-\frac{12\mu Q}{H^2} \alpha(k,\delta)f_p(k,\delta)
\end{equation}
where
\begin{equation}
\label{g}
f_p(k,\delta)=\frac{3}{4}\frac{k^2(1/2-\delta)}{b^{5/2}}\left[\frac{\pi}{2}+\mathrm{atan}\left(\frac{k}{2b^{1/2}}\right)\right]+\frac{3k^3}{8b^2}\frac{1}{(1/2-\delta)}+\frac{1}{(1/2-\delta)}\frac{k}{b}
\end{equation}
and $b=(1/2-\delta)^2-k^2/4$. 
Similarly, the pressure drop across the top gap is $\Delta p_+=\rho g H k - [1-\alpha(k,\delta)]f_p(k,-\delta)$. 
In the lubrication limit (Eq.~\ref{eq:NSL}) the pressure drop across the bottom and top gaps must be equal, $\Delta p_-=\Delta p_+=\Delta p$, and thus the flow split factor is given by
\begin{equation}
\alpha(k,\delta)=\frac{1}{1+f_p(k,\delta)/f_p(k,-\delta)}
\label{alpha}
\end{equation}
Using Eq.~\ref{alpha} the velocity profile for $|x|\leq R$ in the top/bottom $(+/-)$ gap can be cast as 
\begin{equation}
\label{velocity_profile}
u_{\pm}(x,y)=\alpha(k,\mp\delta)\frac{6Q}{h_{\pm}^3}(h_\pm y-y^2),
\end{equation}
and thus the shear stress on the (top/bottom) channel wall is 
\begin{equation}
\label{shear}
\tau_{\pm}(x,y)=\mp \alpha(k,\mp\delta)\frac{6 \mu Q}{h_{\pm}^2}.
\end{equation}
Analytical integration of Eq.~\ref{shear} wall segments above and below the cylinder ($-R\le x \le R$, $y=0$ and $y=H$) gives a shear force (per unit width in the $x$-direction) 
\begin{equation}
F_{s}=-\frac{12 \mu Q}{H} \left[\alpha(k,\delta) f_s(k,\delta) + \alpha(k,-\delta) f_s(k,-\delta)\right]
\label{Fs}
\end{equation}
where
\begin{equation}
f_s(k,\delta)=\frac{k^2}{4b^{3/2}}\left[\frac{\pi}{2}+\mathrm{atan}\left(\frac{k}{2b^{1/2}}\right)\right]+\frac{k}{2b}.
\end{equation}

To find the drag force $D$ (per unit width) on the cylinder for a given flow rate $Q$ we apply a control volume approach. 
Static force equilibrium in the $x$-direction within the channel section containing the cylinder($-R\le x \le R$) gives 
\begin{equation}
\label{drag_total}
[p(-R)-p(R)] H + \rho g (2RH-\pi R^2) + F_{s} + F=0
\end{equation}
where $F=-(D+F_b)$ is the force exerted on the fluid by the cylinder and $F_b=-\rho g \pi R^2$ is the buoyancy force (per unit width).
Introducing Eq.~\ref{pressure_drop} and Eq.~\ref{Fs} into Eq.~\ref{drag_total} gives
\begin{equation}
\label{Drag}
D= \frac{12\mu Q}{H} \left\{ \alpha(k,\delta) \left[f_p(k,\delta) - f_s(k,\delta)\right] + \left[1- \alpha(k,\delta)\right] f_s(k,-\delta) \right\}.
\end{equation}

This analysis further considers the case where the flow rate $Q$ is not prescribed but rather determined by pressure differentials and body forces applied.
When the obstruction to the flow due to the confined cylinder is negligible, which corresponds to the limit of infinitely long channels, the flow rate is 
$Q_{\infty}=[(p_{in}-p_{out})/L + \rho g] H^3 /12\mu$ as predicted for plane Poiseuille flow.
The additional hydraulic resistance caused by the cylinder produces a reduction of the flow expected for plane Poiseuille flow and thus the actual flow rate is $Q=\phi(k,\delta,l) Q_{\infty}$, where $\phi\le1$ is a flow correction factor that needs to be calculated for different confinement configurations and channel aspect ratios. 
Assuming parabolic velocity profiles are recovered at the channel ends,  static equilibrium for the fluid contained in the entire channel gives  
\begin{equation}
(p_{in}-p_{out}) H + \rho g (2LH-\pi R^2) + F_{s} -\frac{12 \mu Q}{H} (L-2R) + F=0,
\label{control-volume}
\end{equation}
where $p_{in}$ and $p_{out}$ are the prescribed pressures at the channel inlet and outlet, respectively.
Combining Eq.~\ref{drag_total} with Eq.~\ref{control-volume} gives the flow rate
\begin{equation}
Q(k,\delta,l)=\frac{(p_{in}-p_{out}) H^2 + \rho g 2LH^2}{12\mu\left[ l + \alpha(k,\delta) f_p(k,\delta) -k \right]}
\label{flow}
\end{equation}
where $l=L/H$ is the dimensionless channel length or longitudinal aspect ratio.
Using Eq.~\ref{flow} for the flow rate we can define the flow correction factor
\begin{equation}
\phi(k,\delta,l)=\frac{Q(k,\delta,l)}{Q_{\infty}}=\frac{l}{\left[ l + \alpha(k,\delta) f_p(k,\delta) -k \right]};
\label{flow}
\end{equation}
as expected for channels that are not fully blocked by the cylinder ($k<1$) we have $\phi\to1$ in the limit $l\to \infty$.
%


%

\end{document}